\documentclass[epjST]{svjour}
\usepackage{graphics}
\begin{document}
\title{Probing the quantum vacuum with ultra intense laser pulses}
\author{B. Manuel Hegelich\inst{1}\fnmsep
\and Gerard Mourou\inst{2} 
\and Johann Rafelski\inst{3} }
\institute{Department of Physics, University of Texas, Austin, TX 78712
 \and IZEST, Ecole Polytechnique, 91128 Palaiseau, France
 \and Department of Physics, University of Arizona, Tucson, AZ 85721\\[0.7cm]}
\abstract{
This article presents: 1) The theoretical background of strong field physics and vacuum structure and stability; 2) The  instrumental developments in the area of pulse lasers and considers the physics case for ultra intense laser facilities; and  3) Discussion of the applied and fundamental uses of ultra-intense lasers. \\[-3.4cm]
Received 13 March 2014/Revieved in final form 24 March 2014\\
Published on line 4 June 2014:\\ 
Eur. Phys. J. Special Topics {\bf 223}, 1093-1104 (2014)\\[2.2cm]
} 
\maketitle

\section{Introduction}
\label{intro}
We survey the use of high intensity laser in exploration of fundamental properties of the vacuum and particle acceleration.
The rapid technological advance in laser pulse technologies creates hope that in foreseeable future it will be possible to reach laser intensities allowing exploration of the empty space, the vacuum, under influence of laser pulse strong fields.  In next generation of high intense laser facilities the peak power will approach the exawatt (EW) regime. That is 100 000 times the world grid power, albeit compressed into a extremly brief instant. This  will help   produce high energy radiation and particles beams with an extremely short time structure offering a new paradigm for the exploration of the structure of vacuum. It will also help  respond to one of the most fundamental questions: how can light propagate in vacuum, how can vacuum define the speed of light and how can it define the mass of all elementary particles. We further present an extended discussion of the current status of high intensity laser pulse production. Particle production in ``empty" space is the historical path which guides our interest and continues to motivate our study of the ``nothing", the vacuum.
\subsection{Vacuum and the \ae ther}
The word ``vacuum" derives from the ancient Greek `distance, interval' and more specifically, the question if the Earth and the Moon are separated by a void, i.e. empty space which we now call vacuum. It is the word ``chaos" which was the prequel for this emptiness. The word chaos, due to its association with creation myths,  mutated to mean the disorder filling the empty space . The word ``aether'' is the pure air breathed by Gods in the Greek mythology, the 5th element (quintessence) of Plato and Aristotle. This `aether' is today `dark energy': we believe that the Universe is  pushed   apart by   the dark  energy  which  accounts for 70\% of all present day energy content in the Universe. This dark energy is not divisible and thus not ponderable, yet its presence is  very tangible. 

The present day context was anticipated by Albert Einstein, who  in his renown Lorentz Lecture of 1920~\cite{1920:Einstein}  stepped  back from his early criticism of the \ae ther, and wrote in conclusion:
{\it Recapitulating, we may say that according to the general theory of
relativity space is endowed with physical qualities; in this sense,
therefore, there exists an \ae ther. According to the general theory
of relativity space without \ae ther is unthinkable; for in such  space
there not only would be no propagation of light, but also no  possibility
of existence for standards of space and time (measuring-rods  and
clocks), nor therefore any space-time intervals in the physical
sense. But this \ae ther may not be thought of as endowed with the
quality characteristic of ponderable media, as consisting of  parts
which may be tracked through time. The idea of motion may not  be
applied to it.} 

Only the original conveys the exact meaning intended by the author. Thus we reproduce   this pivotal paragraph in the original German: {\it 
Zusammenfassend k\"onnen wir sagen: Nach der allgemeinen
Relativit\"atstheorie  ist der Raum mit physikalischen Qualit\"aten
ausgestattet; es existiert also in diesem Sinne ein \"Ather.
Gem\"a\ss\ der allgemeinen Relativit\"atstheorie ist ein Raum
ohne \"Ather undenkbar; denn in einem solchen g\"abe es nicht
nur keine Lichtfortpflanzung, sondern auch keine
Existenzm\"oglichkeit von Ma\ss st\"aben und Uhren,
also auch keine r\"aumlich-zeitlichen Entfernungen im Sinne der
Physik. Dieser \"Ather darf aber nicht mit der f\"ur ponderable
Medien charakteristischen Eigenschaft ausgestattet gedacht werden,
aus durch die Zeit verfolgbaren Teilen zu bestehen; der Bewegungsbegriff
darf auf ihn nicht angewendet werden.}

\subsection{Light propagation in the vacuum}
This was year 1920; quantum physics was not yet discovered.  Einstein could  not yet know  that  the classical space free of matter is filled with quantum fluctuations. Einstein's \ae ther can have structure, being the quantum vacuum, and this vacuum defines, for example,  how light propagates, as noted first  by Werner Heisenberg and his collaborators~\cite{Heisenberg:1935qt,Heisenberg:1936} and Victor Weisskopf~\cite{Weisskopf} . The quantum fluctuations in the vacuum allow light-light scattering and the conversion of electromagnetic field energy into  particle and antiparticle pairs.  

In particular, a light quantum, a photon,  traveling in the vacuum, can fluctuate into a   particle--antiparticle pair. This pair is virtual,  and since the energy of a single light photon is much smaller than that of a material particle pair,  we say, the pair is `off-mass-shall'. This transmutation of a photon into a virtual pair and back, is called `vacuum polarization' since this process also alters  the nature of the Coulomb law at short distance.

If a second photon arrives  just when the first  photon exists in its electron--positron pair state,  it can scatter from  this virtual charged particle pair. In this way,  light scatters  light. In a direct extension of this argument, a strong electromagnetic field applied in the vacuum can deflect the virtual electrons and positrons. Therefore, there is an index of refraction of empty space filled with fields. In principle, light can be bend by an applied electro-magnetic fields, just like in gravity the deformation of space-time geometry  bends light.

In this sense empty space has a structures. Because the Compton wavelength of an electron, $\hbar/mc=386$\,fm (where fermi\,=\,femto meter\,=\, $10^{-15}$ m is the radius of a proton) is a million times shorter than a typical optical wavelength, vacuum structure does not massively obstruct the propagation of light. However, light propagating in the Universe over cosmological distances, in presence of external magnetic fields  experiences nonlinear vacuum effects, such as photon splitting.
For what we think about, it is important to recall that mid-last century  Julian Schwinger  showed explicitly that  a coherent   ideal `plane light wave' cannot scatter from itself, or be influenced by itself, no matter what is the field intensity~\cite{Schwinger:1951nm}.  This is the only form of light known to which  the vacuum is exactly  transparent.

\subsection{Elementary particle mass}
By a natural extension of the argument presented by Einstein, the vacuum not only defines the velocity of light and how light propagates, but it defines the masses of all elementary particles. There are two `standard' particle model  mechanisms of mass generation:
\begin{enumerate}
\item Fermi's weak interactions has been unified with electromagnetic interactions -- the mechanism of this unification includes a new recently discovered particle the Higgs. This particle of mass $M_Hc^2=125$\, GeV (roughly 130 proton masses) comes along with a  large vacuum energy scale $\langle h \rangle =246\,{\rm GeV}\simeq 2M_H$. All elementary particles are thought to derive their mass as a specific fraction of  $\langle h \rangle$. Thus a change in vacuum expectation value  $\langle h \rangle$ of the Higgs field would naturally alter inertial mass of all particles.  
\item The strong interactions provide the actual mechanism explaining the inertial mass of visible matter. The mechanism which confines quarks into a small space domain, and thus assures `color' confinement,  generates via the quantum zero-energy the dominant part of the mass of all matter in our Universe.
\end{enumerate}
In both cases, the physical properties of the \ae ther encode the information which is needed to define the particle mass,   the inertial mass and thus via the equivalence principle also the gravity of visible matter. The vacuum is the carrier of the elementary mass-energy scales. In that sense, \ae ther exists and is in fact the frontier of experimental and theoretical research today. 

\subsection{A short history of the Universe\cite{Rafelski:2013yka}}
The quantum vacuum structure we address  is  present all over the Universe. Let us look  back to the time about 1/100 000 of the second after the Big Bang, when the temperature was  higher than about $kT>kT_{\rm cr}=0.15$\,GeV. We find quarks which can move freely in  the `deconfined' Universe in which nucleons are dissolved and their constituents, quarks, roam freely. At that early time the Universe was nearly symmetric with respect to  quarks and anti-quarks, up to the tiny nano-scale matter-antimatter asymmetry. The residual matter surrounding us is this asymmetry, a remnant from this epoch, after most matter and antimatter annihilated. 

The partners of quarks are  the very light electrons which at that stage of the early Universe already had the properties that we know  today. Contrary to  intuition, the heavy particles, protons and neutrons, turned  out to be more fragile; they dissolve entirely in the quark-gluon phase. The drastic change of one property of the vacuum, the melting of confinement, does not imply that all or any other properties are impacted, much the same  as it is with material systems. 

Well beyond the conditions in which the strong interaction vacuum melts,   at 1000 times higher temperature the electro-weak vacuum will melt and we expect that all visible particles we know loose their mass.  It is believed that at that early time -- scale is a nanosecond --   or shortly before, physical processes were active which lead to the formation of the matter-antimatter asymmetry which sustains our existence. 

Research  into the origin of matter in the Universe relies on ideas that the form of  matter present is dependent on ambient conditions and specifically the temperature, and  that the matter-antimatter asymmetry was created in a different `molten' vacuum state,  present in the early Universe   following on the big-bang. It is commonly believed that even at this matter generating high temperature the dark matter plays mostly a spectator role, considering its weak coupling with visible matter. When exactly the dark matter decoupled from visible matter  is subject to an intense present debate.

\subsection{Recreating the early Universe in the laboratory: Heating the vacuum}
If we are ever to understand the mechanisms of matter creation and matter stability we   need to find a way to    study the vacuum experimentally. The  study of vacuum structure is by necessity focused on experiments involving a small domain of space-time, which follows considering the Stephan-Boltzmann low which allows us to evaluate how much energy is needed to heat a given empty volume to a prescribed temperature.

As temperature rises   we first excite electron and positron pairs, than heavier particles, and at unexpectedly low temperature, with $kT$ being 6 times lower than the energy-mass equivalent of the proton, we   melt the vacuum structure and free quarks replace protons. One must, however, remember that this `low' temperature we talk about  is  far above that  possible even in the center of heavy stars and neutron stars. The lifespan of the hot vacuum in the laboratory is limited by speed of radiative cooling and flow dynamics heat and it is generally accepted that it cannot be longer than the light needs to cross the diameter of this domain.

Particle accelerators offer access to this early Universe quark-gluon plasma\cite{Fromerth:2012fe,Rafelski:2013qeu}.  In collisions of heavy ions at RHIC-BNL, (NY, Long Island) and at CERN (Geneva, Switzerland), the heated strongly interacting matter reaches temperature which `melts' the confining vacuum structure. This is done by accelerating heaviest atomic nuclei to highest energies possible. In most head-on collisions two clouds of 500  quarks which make a heavy atomic nucleus collide, and experiments show that the energy available in the collision is thermalized.  That means that instead of a few high energy particles we observe many thousands newly produced particles which share in the total energy available.

The QGP domain is of the size of an atomic nucleus,  large on scale  of the proton  size, but extremely small compared to our dimensions.  For this reason the energy needed to form QGP  is below 10 erg (6,240 GeV).  The thermal energy content scales with fourth power of the temperature, and third power of radial size. In order to reach the 1000-times higher electro-weak transition temperature  in a volume with 10-times smaller radius,   an energy content which is a billion times higher, that is at the level of a few  kJ, will be required. 

No particle accelerator we can imagine could possibly deliver the required energy to a small volume. If we were to build a solar system sized accelerator the fundamental properties of strong  interactions which are increasingly weakening at high energies (`asymptotic freedom')  limit the amount of usable thermal energy that would be available. However, kJ is well within the energy content available in ultra high intensity pulsed lasers. The technological challenge is here to find a way to focus a good fraction of pulse energy into an elementary particle volume. In this way we could reach energy concentration required for the study of the early Universe physics beyond quark-gluon plasma, that is the electro-weak vacuum structure, baryon stability, and even quintessence, the \ae ther.

On the way to this very extreme concentration of energy a  field mired in physical and technological problems challenge us. One of stepping stones is when we focus  1kJ of energy into radius 0.1\AA. This allows the formation of a near atomic size deconfined quark-gluon plasma phase which has a longer lifespan, allowing the equilibration of electromagnetically interacting  particles with the quarks. Absence of matter brought in by e.g. colliding nuclei results in symmetric  matter -- antimatter   conditions akin to the early Universe.
 
However, while this target radius is 10,000 times larger as in typical accelerator experiments, it is also 10,000 times smaller than typical focal radii achieved by ultrahigh intensity lasers today. The achievable focal radius of a laser is, for a single beam limited by its wavelength, which currently is in the optical/near-IR regime, $\simeq 1\,\mu$m. However, in collisions of many phase synchronized laser beams it is possible to reduce the size of the focal volume well below diffraction limit. We cannot  discuss here in any detail the design of such a futuristic system. However, it will resemble NIF or MegaJoule in that it will involve 100's of concentric beams which will need to be phase locked, a technological challenge that is already today addressed in the context of the ICAN project. 

If such a system can be realized, a plethora of physics becomes available for experimental exploration, and of course we can carry out most interesting experimental research program as the energy density in the focal area increases while the size of the focal area shrinks. For example for rather  `large' nano-size focal domain we still reach   temperatures  100's time greater than in the center of the Sun.  Here we will be able to form in the vacuum  a hot electron-positron-photon plasma. This should allow   exploration of a domain of plasma physics otherwise beyond our reach. Moreover, such a time-pulsed state is   an  intense source of heavier particles, muons and  pions.

\subsection{Vacuum stability and pair production}
The gap between the valence and conduction band of the best insulator, the vacuum, is twice the energy equivalent of the electron mass, $V_0 = 2mc^2  /e =$\,one million Volt. Such high potential differences are commonly achieved in specialized nuclear accelerators (tandems, van der Grafs), however over a rather large distance. The vacuum does not begin to spark since the electron-positron pair must materialize on two ends of the potential well and this is for laboratory devices a macroscopic distance apart. The electric field strength controls the speed of vacuum sparking\cite{Labun:2008re}.

The field strength for which this vacuum decay occurs at  the zepto-second scale (light travels a distance of  the Compton wavelength  in nearly 10 zeptoseconds (Zs), $10^{-21}$\,s) is the so called `Schwinger' critical strength $E_s =1.3 10^{18} $  V/m for which the potential step $V_0$ arises over electron's Compton wavelength.  The critical field is  named after Schwinger, though Heisenberg-Euler have given the result. However Schwinger did discuss the vacuum decay  rate in greater depth, providing the full interpretation of the result  and derived the more general formula applicable to the case when both electric and magnetic fields do not vanish. 

Achieving   this Schwinger field strength $E_s$  is a clearly defined standard  objective of all high intensity laser facilities.  This corresponds to an intensity $I_0=  4.65 \times  10^{29}$ W/cm$^2$. The corresponding laser power when this intensity is considered within the typical  focal domain of  $\mu\mathrm{m}^2$ is $P_0\simeq 4.65 x 10^{21}$ W , that is 4 650 EW  or  4.65 zettawatt (ZW). EW class systems are reaching today experimental design stage, as we discuss below. An approach to reach ZW regime is described below. For comparison, one can buy off-the shelf a PW class (P=peta$=10^{15}$) laser system.

Following the work of Heisenberg and Schwinger of 60-70 years  ago we know (but often ignore) that any macroscopic electric field is a metastable state capable torapidly decay  into particle pairs. Akin to the radioactive decay, the field materialization to pairs process involves tunneling, and thus as the macroscopic field decreases, the lifespan increases rather rapidly. For a field of strength $E_s/30 = 5\, 10^{16}$ V/m the lifespan is at the level of the lifespan of the Universe $\simeq 10^{21}$s\cite{Labun:2008re}, which means that for all practical purposes such laboratory field configurations are stable.

For fields closer to the Schwinger value $E_s$ we can observe massive materialization of pairs for fields existent at time scale of fs to as ($10^{-15}$--$10^{-18}$\,s). The materialization of electrical fields into electron-positron pairs is a diagnostic tool allowing us to reach an  understanding how well we achieved the best laser  energy focus. The abundant formation of electron-positron pairs is the first of many vacuum effects we will be looking for as we strive to focus the laser energy into smaller and smaller volumes.

\section{Pulsed lasers}
\subsection{The making of the ultra high intensity lasers: CPA}
All ultrahigh intensity lasers (UHIL) systems are pulsed lasers. Instead of emitting a continuous wave (cw) of light, like e.g. a presentation laser pointer, they emit pulses of light of short duration. Recapitulating, there are three principle parameters to consider when discussing ultra-intense lasers: the total pulse energy $E_\mathrm{pulse}$ , the pulse duration $t_\mathrm{pulse}$ and the achievable focal size $F=\pi R_\mathrm{focus}^2$,  where the light wavelength limits the focal radius $R_\mathrm{focus}$. From these three parameters then follows the achieved intensity:  $I = E_\mathrm{pulse} /(t_\mathrm{pulse} \cdot  F)$. According to Maxwell equations, the laser intensity is proportional to the square of the field strength present in the  pulse.

This field strength in the pulse is responsible for vacuum instability and thus we are interested in increasing $I$. This can be achieved either by increasing the pulse energy, or decreasing the pulse duration and decreasing the size of the focal spot. The spot size for one or a few pulses is  limited  by the laser wavelength. The laser energy and not the power  determines the laser size and cost, which entails to work with the lowest pulse energy possible. Therefore, the most economical and effective way to reach high peak power pulses would be to amplify the shortest possible pulse. However,   as the pulse gets amplified, nonlinear effects due  to the large intensity set in. Notably, the index of refraction $n$ becomes a linear function of $I$. This effect will: 1) alter the laser  beam quality, making impossible to focus the beam on a diffraction limited spot size;  2) traversing  any component, the beam may coalesce into small filaments, whose large intensity may irreversibly damage the materials. So, since the mid-sixties, efficient, ultrashort pulse amplification that would preserve the beam quality,  seemed unattainable. However the solution was offered in 1985 with the concept of Chirped Pulse Amplification (CPA) invented by one of us~\cite{CPA}. 

In CPA, a short pulse is first produced by an oscillator. Its energy is at the nJ level. It is stretched in time by a very large factor up to $10^6$ by dispersive elements, such as a pair of diffraction gratings.  This is possible because of the large amount of Fourier frequencies forming the ultrashort pulse. Each frequency  takes a different route, and will take  a different time to traverse the dispersive element. Once the frequency components have been dispersed the pulse has been stretched, with the red part of the spectrum ahead followed by the blue part.  Note that this operation has not significantly changed the pulse energy content. Consequently, the pulse intensity has dropped in the same ratio, e.g. $10^6$, implying that it could be amplified to an energy level $10^6$   times higher and still remain compatible with the available optical conditions. Because of the low intensity level, the beam quality and the laser component integrity will be preserved.  Once the pulse is amplified, it is re-compressed to its initial time structure  using a compressor that will exactly undo what had been done by the stretcher. The compressor  will reroute the frequencies in such a way that they will all  travel throughout the entire system an identical optical length.  We can marvel at how well this concept works, if we consider that in modern CPA, the pulse is stretched and compressed by a factor 106, in-between amplified by a very large factor.

With CPA,  the intense and ultra intense lasers  performed a quantum leap in achieving shorter pulse durations and therefore higher power and  higher intensities. At the moment there are two types of amplifying media: \begin{enumerate}
\item Nd:Glass  as laser medium, with pulse durations of ps and pulse energies in the 100 to 1000J range and a shot rate  of  about one pulse an hour. The first PW system, the now decommissioned Nova PW at LLNL belongs to this class, as does the Vulcan PW (RAL, UK), Gekko PW(ILE, Japan), Trident (LANL, USA), Titan (LLNL, USA), the Texas PW (UTA, USA) and Omega EP (LLE, USA). 
\item Ti:Sapphire as laser medium. These systems amplify shorter pulse on the order of tens of fs and correspondingly smaller pulse energies of few Joules, with a shot rate of 10 Hz. 10-100 TW systems of these types are widespread and already commercially available today. Examples of those systems approaching the PW level are the just commissioned Gemini laser (RAL, UK) and the PW system at the Shanghai Institute of Optics and Fine Mechanics (SIOM, China) a PW system at the University of Nebraska, Lincoln (USA) and the PW LUIRE at the ILE(France).   All these systems reach  intensities on target of $I=10^{19}$--$10^{21}$ W/cm$^2$, with the highest published intensity being claimed by the Hercules laser at the University of Ann Arbor (USA), with $10^{22}$ W/cm$^2$ which involves a pulse of 10J, lasting 30fs.
\end{enumerate}

\subsection{OPCPA -- Optical parametric chirped pulse amplifier}
Few yeas after the advent of CPA an elegant embodiment was demonstrated in which the conventional laser amplifier was replaced by an optical parametric amplifier(OPA). The OPA is an amplification mechanism that relies on nonlinear processes, where photons called pump are broken into photons called signal and idler. Among the advantages we note, the very high gain per pass, the very large bandwidth and the fact that thermal effects are absent in the amplifying crystal. 

But, there are not only pros, among the cons we note: 1) OPCPA needs an excellent pump beam; 2) OPCPA does not store energy so requires a precise temporal overlap of the pump  and signal pulse and so the pump needs to be in the single nanosecond range, which is quite short; 3) the efficiency is about half  of Ti:sapphire CPA; 4) as a corollary of 2) and 3), the repetition rate will be inherently lower than for a straight CPA. These considerations become dominant  for a large scale laser system, where the cost of the system scales almost linearly with the laser energy.

The concept of OPCPA (Optical Parametric Chirped Pulse Amplification) was proposed and demonstrated by Piskarskas et al~\cite{OPCPA}. On large systems it was pioneered at RAL. Taking the OPCPA approach to its next level in pulse energy, the Institute for Applied Physics in Nizhny-Novgorod, Russia demonstrated a 0.6 PW, 30J, 40fs system.
One of the highest power systems operating today, the Texas PW laser uses a hybrid of these technologies. It uses OPCPA technology to achieve most of its gain at high bandwidth, up to a pulse energy level of $\simeq 1$\,J, which then gets further amplified in a set of mixed Silicate/Phosphate glass amplifiers up to its final energy level of about 160\,J. At a pulse duration of less than 150\,fs, that equates to slightly more than 1\,PW. Currently, the laser can be focused to intensities of $\simeq 10^{21}$ W/cm$^2$. Upgrades currently underway will add large aperture adaptive mirrors, short focal length focusing mirrors and a prepulse reduction system which will enable high contrast laser target interactions at $\simeq 5\times 10^{22}$ W/cm$^2$ over a duration of ~50 optical cycles enabling the highest integrated intensity interactions on any current system and for the next few years.  Larger systems  are on the drawing board and are currently being designed.

\subsection{The highest intensity laser planned today will provide intensities $\simeq 10^{26}$\,W/cm$^2$ } 
In order to reach the highest intensity while maintaining a good efficiency and a high repetition rate, the architecture of a laser today must be an hybrid-OPCPA-CPA system. It will utilize the OPCPA as the front end, where the efficiency is less of an issue, to preserve the pulse bandwidth to the highest energy level possible. However, because the system efficiency is dictated by the last amplifier, the latter will be Ti:sapphire CPA. This material possesses the largest known amplifying bandwidth.  This architecture will ensure simultaneously the shortest pulses, i.e. $\simeq 10$\,fs, the  highest energy per pulse(few kJ), the highest overall efficiency and the highest repetition rate. The final peak power should be in the range of 200\,PW corresponding to an intensity, greater than $10^{25}$\,W/cm$^2$ for a diffraction limited spot size at 800nm. 

\subsection{Bridging the gap; all the way  to the Schwinger intensity regime}
Note that this phenomenal intensity, $10^{26}$\,W/cm$^2$,  is still 3 orders of magnitude below the Schwinger field instability limit.Simulations show that at this level few if any pairs will  be created. There is need to boost the intensity all the way to the Schwinger limit.  Heretofore, we envision three ways of bridging the intensity gap: 
\begin{enumerate} 
\item
{\it  Relativistic compression.} Pulse compression, at the attosecond
(10-18) or zeptosecond (10-21)level, could offer the possibility of also producing intensities in the Schwinger limit. Naumova et al,\cite{Nau} described one embodiment resulting from the relativistic motion of the overdense plasma critical surface. Simulations show  that the critical surface moving at the speed of light could compress each cycle by a factor as large as several thousands with good efficiency. In addition, the extraordinarily large light pressure could bow the critical surface to make a spherical mirror  that will focus by reflection the light on a tighter spot to reach the Schwinger intensity regime.
\item
{\it Lorentz-boosted field.}
It is possible to produce high energy electron beams with a very high Lorentz $\gamma$ factor. For example consider  a  50GeV beam ($\gamma=10^5$). For a  laser beam propagating against the electron beam, its field in the frame of reference of the electrons, will be enhanced by  $ \gamma=10^5$   This  should bring the field acting in the rest frame of the electron to a level much higher than the critical intensity\cite{Rafelski:2009fi,Hadad:2010mt,Labun:2010wf}. 
\item
{\it Ultra-short pulse.} Finally, we want to mention a path which is more hypothetical but that could be envisioned in the next 10-20 years. It is the direct generation of ZW ($10^{21}$\,W) pulses. This will require the production of 10\,fs duration pulse at the MJ level. It could be conceptually performed with ``The Laser MegaJoule", built in France or the NIF in the USA. The few MJ at $2\omega$ available by these lasers could be used to pump a Ti:sapphire matrix of approximately 10\,m diameter. We would use an OPCPA front end to preserve the short pulse bandwidth. A 10\,fs pulse with MJ energy, would lead to a zettawatt class pulse. It could be focused by a 10\,m, segmented telescope mirror to a wavelength diameter spot size to produce Schwinger level intensity. A preliminary description of this ambitious system is discussed by  Mourou and Tajima\cite{TM}. 
\end{enumerate}

\begin{figure}
\resizebox{1.02\columnwidth}{!}{\includegraphics{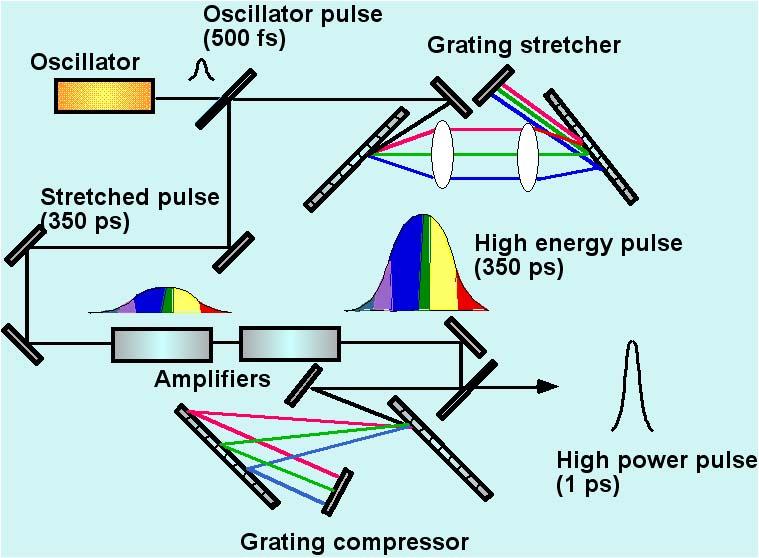}}
\caption{Ultra short pulses cannot be amplified directly. Even with a modest amount of energy, the intensity in the amplifier can reach phenomenal values that will produce undesirable nonlinear effects, namely, wave front distortions, component damages, etc.   To decrease the intensity without changing neither the input pulse energy -- necessary for energy extraction -- nor the pulse bandwidth, we stretch the pulse with dispersive elements by say $10^3$ to $10^5$ times decreasing the intensity by the same ratio. Once the pulse is stretched it can be amplified by a factor say  $10^6$ to $10^{12}$. Once the pulse is amplified it will be recompressed by a dispersive devices (pair of gratings) to the pulse initial value. Note that the sum of the phase functions of the stretcher, the compressor and the amplifier optical elements must be equal to zero over the full bandwidth. The amplifiers can be a conventional laser amplifier or an optical parametric amplifier(OPCPA) or both.\label{fig:1}}
\end{figure}

%

\section{Use of ultra intense lasers}
\subsection{Particle acceleration}
The extraordinarily rapid advance in laser technology is driven by the promise of compact and cheap particle accelerators. As predicted in their seminal paper by Dawson and Tajima~\cite{Tajima:1979bn}, in the relativistic laser pulse regime, electrons get accelerated to the speed of light within a fraction of the laser pulse, making the $\vec v \times \vec B$ term of the Lorentz force non-negligible and resulting in a net forward acceleration. Depending on the exact conditions of the target, like plasma density, temperature, scale length and ionization degree, various schemes for electron acceleration, such as direct laser acceleration, wake-field acceleration, bubble acceleration, etc. can be exploited as discussed in other contributions to this volume and in Ref.~\cite{Ledingham:2010zz}. 

This has led to the demonstration of GeV electron beams over acceleration distances of mere mm as compared to hundreds of meters in conventional accelerators. This opens the possibilities of utilizing such beams  heretofore were only accessible at large accelerator facilities and light sources at the table-top or laboratory scale and would therefore allow their application in Universities, hospitals or industry. Especially the possibility of a table-top x-ray free electron lasers (XFEL) as a 5th generation light source for material science research, bio-molecular imaging and fundamental research as well as medical diagnostic purposes as a source for ultrahigh contrast, low dose medical imaging.

Also  laser accelerated electrons can be used to act as a mediator to in turn accelerate heavier particles such as protons, carbon ions or other species. While electron acceleration schemes usually use low density gas targets to optimize the electron energy, in ion acceleration one needs to optimize the electron number at ~MeV energies. To this purpose, solid density targets, usually micron sized metal foils are employed. The laser, when interacting with those targets, accelerates electrons which penetrate the foil and setup a quasi static electric field at the rear side, forming a virtual cathode. This field, which is of the same order of magnitude that the laser field, i.e. many TV/m, ionizes and accelerates the ions of the rear surface. This process, known as Target Normal Sheath Acceleration (TNSA), has demonstrated MeV ion acceleration of micron distances, again outclassing conventional accelerators by many orders of magnitude in terms of shear acceleration gradient. 

Recently  a new ion acceleration mechanisms, dubbed Break-Out Afterburner Acceleration (BOA) \cite{H1} has been observed in simulations~\cite{H2,H3,H4,H5}, and in experiments~\cite{H6,H7,H8,H9,H10,H11,H12,H13} , demonstrating ion energies of greater 100 MeV/nucleon at intensities $10^{20}$--$10^{21}$ W/cm$^{2}$ , specifically 160 MeV protons and 1 GeV carbon ions from a 130 TW laser were obtained. This portrays the potential of increasing the ion energies into the GeV region for currently achievable intensities of $10^{21}$ W/cm$^2$ and well beyond that for higher intensities. 

The BOA mechanism works via a relativistic, kinetic plasma instability, and requires use of ultra thin, nanometer scale targets for current laser parameters. In order for the laser to interact with such an extraordinarily thin target, a fraction of its own wavelength in size, the contrast of the laser pulse, i.e. the ratio of the foot of the pulse to its peak, has to be controlled within ~12 orders of magnitude for today's ultrahigh intensity lasers. That corresponds to controlling the distance Earth-Moon to within the fraction of the thickness of a hair, which illustrates the technical challenge underlying the experiments.

Recent experiments at the Los Alamos Trident laser managed to achieved just that, accelerating Carbon ions to 0.5 GeV using a laser intensity of $\simeq 10^{20}$ W/cm$^2$ and demonstrating BOA acceleration for the first time and reaching energies within an order of magnitude of those required e.g. for cancer therapy. If the technological issues of repetition rate, shot-to-shot fidelity, etc  are solved by future improvements in laser technology this will than enable to transfer heavy ion therapy from a few large synchrotron facilities to a much smaller and cheaper laser system suitable tolarge hospitals, making a promising new therapy available to a wider public.

\subsection{Brilliant X-ray sources}
Recent numeric simulations show that even the thinnest possible target foils, only nanometer thick, are capable of absorbing laser pulse energy at percentile level. If the laser energy conversion into  kinetic energy of matter is the objective one can easily envisage a relatively good conversion efficiency: the    laser amplification process is by necessity made efficient in order to allow high shot repetition rate, thus the overall efficiency to particle beam is expected to be in 10\% range.

Simulations also show that these nm targets required for BOA might be used as sources for ultra dense, mono-energetic electron bunches, which via Compton scattering could be used as the basis for an ultra compact, ultra brilliant coherent x-ray source, surpassing even other laser-driven schemes currently under consideration.  However,  these examples of applications are still far away from technical realization  and require much more and dedicated research, they amply illustrate the potential societal payoff from fundamental research into ultrahigh intensity lasers.

\subsection{Strong fields}
A side effect of particle acceleration directly relevant in the strong field context is the separation of electrons from ions in laser-foil interactions. Experimental simulations of this situation have revealed that the electrical field strength which develops between electrons and the trailing ions is at the level of the fields present in the laser pulse. This is of importance since we are interested in studying the behavior of the vacuum in presence of strong fields, and this mechanism is allowing to convert the transverse laser field into the field string.

Similarly, when we consider interaction of circularly polarized laser pulse with a foil, there is on the surface of the pulse a circular motion, a current induced in the foil which continues into the vacuum. This also generates a strong solenoidal magnetic field. The electrons which are in circular motion around the  propagation emit intense synchrotron radiation which can be of considerable importance to many applications.

\subsection{Heavy particle production}
Finally, we must look at the electron-ion interaction in the foil. We can  optimize   electron-nucleus interactions  to yield hadrons such as pions, antinucleons,  or muons.  Thus the laser-foil interaction region could serve   as source of secondary particles. Of course for applications one would need to master repetition rates at scale of Hertz. However, a non-negligible advantage of this scheme is the pulsed nature of the source and its extraordinary intensity.  

\subsection{\AE ther and quantum vacuum}
We move here back from the more applied interactions with matter to the laser-vacuum interaction. At issue is the question what exactly are the holes we ``burn" in the vacuum. Einstein viewed  a local  change of  the a \ae ther to be in conflict with the philosophical considerations requiring the uniqueness of the laws of physics in the entire Universe. In his 1920 essay  Albert Einstein did not allow  ``pieces" of the \ae ther   with a modified   structure. He says:
         {\it The special theory of relativity forbids us to
         assume the \ae ther consist of particles observable through time,
         but the hypothesis of \ae ther in itself is not in conflict with the
         special theory of relativity. Only we must be on our guard against
         ascribing a state of motion to the \ae ther.} This text in original reads: {\it  
Das spezielle Relativit{\"a}tsprinzip verbietet uns, den {\"A}ther 
als aus zeitlich verfolgbaren Teilchen bestehend anzunehmen,
aber die {\"A}therhypothese an sich widerstreitet der speziellen
Relativit{\"a}tstheorie nicht. Nur mu\ss\ man sich davor h{\"u}ten,
dem {\"A}ther einen Bewegungszustand zuzusprechen.}
Indeed, a vacuum which can be torn into pieces behaves just like ponderable matter,   an excited hot vacuum domain hole at first sight seems to be just another particle.  The present day study  of domains of quantum vacuum with different local structure, the quark-gluon plasma,  clearly  departs from   Einstein's point of view. In departing from Einstein views  we view the quantum vacuum as a wider concept  with its properties rooted in quantum physics and  reaching  beyond the \ae ther.

%

\end{document}